\documentstyle[aps,preprint,amsfonts,epsf]{revtex}
\begin{document}
\def\dfrac#1#2{{\displaystyle{#1\over#2}}}
\preprint{LA-UR-}


\title{
Impurities and Conductivity\\ in a $\protect\bbox{D}$-wave
Superconductor\footnotemark} \footnotetext {Paper presented at the Los
Alamos Conference on Strongly Correlated Electron Systems, December
1993}
\author{
A.\ V.\ Balatsky\protect\cite{ab}
}
\address{
Theoretical Division, Los Alamos National Laboratory, Los Alamos, NM
87544}

\maketitle

\begin{abstract}
Impurity scattering in the unitary limit produces low energy
quasiparticles with anisotropic spectrum in a two-dimensional $d$-wave
superconductor. We describe a new {\em quasi-one-dimensional } limit
of the quasiparticle scattering, which might occur in a superconductor
with short coherence length and with {\em finite} impurity potential
range. The dc conductivity in a $d$-wave superconductor is predicted
to be proportional to the normal state scattering rate and is
impurity-{\em dependent}. We show that {\em quasi-one-dimensional }
regime might occur in high-$T_c$ superconductors with Zn impurities at
low temperatures $T\lesssim 10$~K.

\end{abstract}



In this short note I will address {\em the role of a strongly
scattering impurities with finite range on the dc conductivity} in a
short coherence length superconductor. This is report on the work,
done in collaboration with A. Rosengren and B. Altshuler \cite{BRA}.

It is well known that scalar impurities are pair breakers in $d$-wave
and any other nontrivial pairing state
superconductor~\cite{Rice,Gorkov,Lee}. They produce a finite lifetime
of the quasiparticles in the nodes of the gap, a finite density of
states at low energy, and a {\em finite} low frequency conductivity at
low temperatures, ignoring localization effects. For the special case
of a 2D superconductor with a $d$-wave gap, a straightforward
calculation yields the surprising result that {\em dc} conductivity
$\sigma(\omega\to 0)$ is a ``universal'' number~\cite{Lee}, {\em
independent} of the lifetime of quasiparticle (but dependent on the
anisotropy ratio of the velocities of the quasiparticle in the node of
the gap)~\cite{Fradkin}. However, recent experiments on microwave
absorption in YBCO crystals with Zn impurities~\cite{Bonn} show a {\em
linear} temperature dependence of the conductivity for pure samples,
evolving to the quadratic behavior for higher impurity concentration,
and low-temperature conductivity, inversely proportional to the
impurity concentration.

 i)~We find a new {\em quasi-one-dimensional regime} for dc
conductivity in superconductors with a short coherence length $\xi$,
comparable to the range of impurity potential $\lambda$. The
quasiparticle contribution to {\em dc } conductivity is governed by
self-energy $\Sigma(\omega\to0)=-i\gamma$ and by the phase space
available for low-energy quasiparticles. The quasiparticle dispersion
is strongly anisotropic in the vicinity of the nodes in a 2D $d$-wave
superconductor: $E_k=\sqrt{v_1^2k_1^2+v_F^2k_3^2}$ and
$v_1/v_F\sim\Delta_0/\epsilon_F$. Here we linearized spectrum in the
vicinity of the nodal point close to $({\pi\over{2}},{\pi\over{2}})$,
so that $k_1$ is the momentum along the Fermi surface and $k_3$ is
perpendicular. We find that the overall contribution to the
conductivity depends on the ratio of the energy of the quasiparticle
to the scattering rate $v_1\lambda^{-1}/\gamma$, $\lambda$ is the
range of the impurity potential, and we get at $T=0$:
\begin{equation}
\sigma(\omega\to 0)=\frac{e^2}{2\pi\hbar}~\frac{2}{\pi^2}~
\frac{v_F}{v_1}~\left(1+\left(\frac{\gamma}
{2v_1\lambda^{-1}}\right)^2\right)^{-1/2}.
\label{sigv}
\end{equation}
For $v_1\lambda^{-1}/\gamma\ll 1$ quasiparticle dynamics is
essentially {\em quasi-one-dimensional } and conductivity {\em
depends} on the impurity concentration $\sigma_{Q1D} \sim
n^{-1}_{imp}$. Our model predicts that the dc conductivity at low
temperature should be {\em proportional to the scattering rate in the
normal state}. This limit might occur in high-$T_c$ superconductors,
for which we estimate $\lambda/a \sim 1-3$ and
$\Delta_0/\epsilon_F\sim 10^{-1}$. In the limit $\lambda\to 0$
Eq.~(\ref{sigv}) gives the ``universal'' dc conductivity $\sigma_{2D}
\sim v_F/v_1$, found in~\cite{Lee}.

To explain this effective {\em change of dimensionality} we note that
transverse momentum is limited by $k_1<2/\lambda$ and quasiparticle
dispersion on such a small scale is irrelevant, compared to $\gamma$.
The condition for this to occur is precisely $v_1 \lambda^{-1}/\gamma
\ll 1$. The transverse (along the Fermi surface) scattering does not
contribute effectively to the conductivity; we call this case a {\em
quasi-one-dimensional } limit. The existence of this limit is the
result of the {\em finite} impurity range $\lambda$~\cite{Born}. In
the opposite limit $v_1 \lambda^{-1}/\gamma \gg 1$, which holds for
``zero" impurity range, we recover standart unitary scattering results
\cite{Hirschfeld,Doug2,Gold}.

 ii)~Here we will explain the assumptions we made to calculate
conductivity. We assume that impurities are strong scatterers with
s-wave phase shift $\delta_0({\bf q}) \simeq \pi/2$, for $|{\bf q}| <
\lambda^{-1}$, where ${\bf q}$ is wavevector, counted from the Fermi
wavevector. This assumption is well supported by experiments on
cuprates with Zn imputrities.  The origin of strong potential impurity
scattering in high-$T_c$ superconductors is the highly correlated
antiferromagnetic nature of the normal state. The second assumption is
about the finite range $\lambda$ of the impurity potential, which
palys role of the momentum cut off in the momentum dependence of phase
shift $\delta_0({\bf q})$. It is as well motivated by the fact that
high-$T_c$ superconductors have a substantial antiferromagnetic
coherence length $\xi_{AFM}\sim 3a$ at the transition temperature. A
scalar impurity will produce distortions in magnetic correlations on
the range of the $\xi_{AFM}$. On the other hand superconducting
coherence length $\xi\sim 20$~{\AA} is comparable to this scale and
thus, the range of the potential is finite on the scale relevant for
superconductivity.  This point should be contrasted to the case of
heavy-fermion superconductors, where the coherence length is $\sim
10^3$~{\AA}, and therefore, any potential impurity will have its range
substantially shorter than the coherence length.  We retain this cut
off finite and on the order of few lattice constants ($\lambda \sim
2a$). This implies that impurities still are well screened and s-wave
scattering is dominant.

iii)~To calculate the quasiparticle conductivity we use  lowest
order bubble diagram with self-consistent Green functions with {\em no
vertex corrections}, see for example~\cite{Lee}. For
the  dc  conductivity we get~\cite{Mahan}:
\begin{eqnarray}
\sigma(\omega\rightarrow0)=\frac{e^2}{\hbar}~
\frac{4 v^2_F}{\pi^2}\mathop{{\sum}'}_{\bbox{k}}\int
d\epsilon~(-\partial_{\epsilon}n(\epsilon))
(|G''(\bbox{k},~\epsilon)|^2+|F''(\bbox{k},~\epsilon)|^2),
\label{cond}
\end{eqnarray}
where, linearizing  quasiparticle spectrum in the vicinity of
nodes,
$G''(\bbox{k},~\omega=0)=\gamma/(\gamma^2+(v_1k_1)^2+(v_Fk_3)^2),~F''(\bbox{k},~\omega
=0)=0$. The momentum integral in Eq.~(\ref{cond}) is cut off at $|{\bf k}| \leq
2/\lambda$ and it  yields the final
formula Eq.~(\ref{sigv}) for $T=0$ with $O(T^2)$ corrections.

For the particular case of strong disorder $v_1 \lambda^{-1}/\gamma
\ll 1$, considered in \cite{BRA}, relation between scattering rate in
the superconducting state $\gamma = i\Sigma(\omega\to0)$ and
scattering rate in the normal state $\Gamma = n_{imp}/\pi N_0$ is
$\gamma = \pi/8 \ p_F \lambda \ \Gamma$. Note that the scattering rate
at low temperatures is {\em linearly} proportional to $\Gamma$, as
opposed to the $\Gamma^{1\over2}$ dependence in the standart unitary
scattering case for $v_1 \lambda^{-1}/\gamma \gg 1$. The assumption
$v_1 \lambda^{-1}/\gamma \ll 1$ is consistent at $\lambda \sim 2a$ for
$\Gamma \geq 20 K$. This estimate shows that the {\em
quasi-one-dimensional} regime of quasiparticle scattering should occur
in not too clean samples at $T < \gamma$. In this limit scattering
rate in the superconducting state is of the same order as the normal
state scattering rate $\gamma\sim 2\Gamma\sim 40$~K for
$p_F\lambda\sim 6$ and is similar to the scattering rate in the {\em
2D} limit: $\gamma/\tilde{\gamma}\sim\sqrt{\Gamma/\Delta_0}~p_F\lambda
\simeq 1$.  The finite density of states $N(\omega\rightarrow0)/N_0 =
\Gamma/\Delta_0 \sim n_{imp}$, {\em linear} in  impurity
concentration, is generated  as well.

Using the above estimates we find the conductivity
in {\em quasi-one-dimensional} regime
\begin{eqnarray}
\sigma(\omega\rightarrow0)=\frac{e^2}{\pi\hbar}~\frac{16}{\pi^3}~
\frac{\hbar}{m\lambda^2\Gamma}.
\end{eqnarray}
It is smaller than the normal state conductivity
$\sigma(\omega\rightarrow0)_{\rm
normal}=(e^2/\pi\hbar)~(\epsilon_F/\Gamma)$ due to small factor
$\hbar/(m\lambda^2\epsilon_F)\lesssim 1$ with $\lambda >a$.
Conductivity is also impurity-{\em dependent},
$\sigma_{Q1D}\sim\Gamma^{-1}\sim n^{-1}_{imp}$. This model predicts
that the dc $\sigma$ at low temperatures should be {\em inversly
proportional to the scattering rate in the normal state and to the
impurity concentration}. We emphasize that {\em both} a higher value
of the conductivity in the superconducting state {\em as well as}
strong impurity dependence at low temperatures are observed
experimentally in microwave absorption of YBCO~\cite{Bonn}.  It should
be pointed out that we are interested in only elastic scattering and
strong inelastic contribution to the scattering rate above $T_c$ is
not considered in this model.

iv)~One can be almost certain that for dirty enough superconductors
the {\em quasi-one-dimensional} regime will occur, since impurities
will lead eventually to a very high scattering rate. The question
remains about the competing phenomena, such as the localization of
quasiparticles, which might occur earlier than {\em
quasi-one-dimensional} regime. It is interesting to apply this model
to the Zn impurities in the YBCO.  The applicability of the results
presented here to the different high-$T_c$ materials depends on the
ratio of scattering rate to the relevant quasiparticle energy. For the
same impurity concentration different cuprates may be in different
regimes, depending on $\xi_{AFM}$, $\Delta_0/\epsilon_F$, and
$\Gamma$.

\underline{Acknowledgements} I would like  to thank  A. Rosengren
 and B. Altshuler for collaboration and discussions.
 I am grateful to E. Abrahams, P.\ Lee, P.\ Littlewood,
D.\ Pines and D.\ J.\ Scalapino for discussions.


\begin{references}
\bibitem{Born}
The fact that impurity scattering is {\em unitary} is crucial for this
{\em quasi-one-dimensional } limit to occur. Indeed the simple estimate
shows that for the {\em Born} limit with weak scattering
$v_1\lambda^{-1}/\gamma\gg 1$.
\end{references}
\end{document}